\documentclass[12pt,preprint]{aastex}
\usepackage{natbib}

\shorttitle{The Millisecond Pulsars in NGC~6760}
\shortauthors{Freire et al.}

\begin{document}

\title{The Millisecond Pulsars in NGC~6760}

\author{Paulo C. C. Freire\altaffilmark{1},
  Jason W. T. Hessels\altaffilmark{2},
  David J. Nice\altaffilmark{3},
  Scott M. Ransom\altaffilmark{2,4},
  Duncan R. Lorimer\altaffilmark{5} \&
  Ingrid H. Stairs\altaffilmark{6}}
\altaffiltext{1}{NAIC, Arecibo Observatory, HC3 Box 53995, PR 00612,
  USA; {\tt pfreire@naic.edu}}
\altaffiltext{2}{Department of Physics, Rutherford Physics Building,
  McGill University, 3600 University Street, Montreal, Quebec, H3A
  2T8, Canada; {\tt hessels@hep.physics.mcgill.ca}}
\altaffiltext{3}{Physics Department, Princeton University, Box 708,
  Princeton, NJ 08544; {\tt dnice@princeton.edu}}
\altaffiltext{4}{National Radio Astronomy Observatory, 520 Edgemont
  Rd., Charlottesville, VA 22903 USA; {\tt sransom@nrao.edu}}
\altaffiltext{5}{University of Manchester, Jodrell Bank Observatory,
Macclesfield, Cheshire, SK11 9DL, UK; {\tt drl@jb.man.ac.uk}}
\altaffiltext{6}{Department of Physics and Astronomy
University of British Columbia, 6224 Agricultural Road,
Vancouver, BC V6T 1Z1 Canada; {\tt istairs@astro.ubc.ca}}

% \email{pfreire@naic.edu}

\begin{abstract}
We present the results of recent Arecibo and Green Bank observations
of the globular cluster NGC~6760.  Using Arecibo, a phase-coherent
timing solution has been obtained for the previously known binary
pulsar in this cluster, PSR~J1911+0102A. We have also discovered a new
millisecond pulsar in NGC~6760, PSR~J1911+0101B,
an isolated object with a rotational period of 5.38\,ms and a
dispersion measure ${\rm DM} = 196.7\,\rm cm^{-3}\,pc$. Both pulsars are
located within 1.3 core radii of the cluster center and have negative
period derivatives. The resulting lower limits for the accelerations
of the pulsars are within the range expected given a simple model of
the cluster. A search for eclipses in the PSR J1911+0102A binary system
using both telescopes yielded negative results. The
corresponding limits on the extra gas column density at superior
conjunction are consistent with the hypothesis that the observational
properties of ultra-low-mass binary pulsars like PSR~J1911+0102A are
strongly affected by the inclination of the orbital plane of the
system. Among globular cluster pulsar populations, that of NGC~6760
exhibits one of the largest known spreads in DM. This quantity
seems to be roughly proportional to a cluster's central
DM; this suggests that the observed spread is caused by a turbulent
interstellar medium at spatial scales of 1 pc.
\end{abstract}


\section{Introduction}\label{sec:intro}

NGC~6760 lies towards the inner Galaxy; $l\,=\,36.11^\circ,
b\,=\,-3.92^\circ$ \cite{dm93a}. It is one of the metal-rich Galactic
globular clusters (GCs) studied by Heitsch \& Richtler
(1999)\nocite{hr99}. Apart from its metallicity, it is quite an ordinary,
moderately bright cluster (total absolute visual magnitude $-$7.86) and not
especially condensed: $c\,\equiv\,\log (\theta_t / \theta_c)
\,=\,1.59$, where $\theta_c\,=\,0.33\arcmin$ is the angular core
radius\cite{har96}\footnote{For an updated list of the GC
parameters presented in this article, and for the related
bibliographical references, see
\url{http://www.physics.mcmaster.ca/resources/globular.html}} and $\theta_t$
is the angular tidal radius. The estimated distance, taking into
account differential reddening, is $D=9.5$~kpc \cite{hr99}. Not taking
that effect into account, one obtains a distance of 7.4~kpc, as
in Harris (1996).

PSR~B1908+00, now known as PSR~J1911+0102A (henceforth PSR~A),
was discovered in a
previous 1400-MHz Arecibo search of NGC~6760 \cite{dma+93}. The pulsar
is a member of a binary system, and its orbital parameters are
remarkable in several respects. The orbital period was, at that time,
one of the shortest known for a binary pulsar (3.38~h), while the
companion mass was the lowest known: only 0.02~M$_{\odot}$, assuming a
pulsar mass of 1.4~M$_{\odot}$ and an orbital inclination angle $i=60^\circ$.
Although the orbital parameters of PSR~A closely resemble
those of the original eclipsing pulsar, B1957+20 (Fruchter, Stinebring
and Taylor, 1988),\nocite{fst88} no eclipses were observed in this new
binary system, despite its shorter orbital period.

PSR~A is now one of the eleven very low mass binary pulsars (VLMBPs) known
in globular clusters. These binaries have mass functions smaller than $3
\times 10^{-5} \rm M_{\odot}$; see Freire (2004)\nocite{fre04} for a
recent review of their properties. Analyzing the properties of the large
VLMBP sample in 47~Tuc, Freire~et~al.~(2003) noted that the occurrence of
eclipses among these systems is strongly correlated with the mass
function of the system. This can be understood if the range of
secondary masses ($m_c$) in these systems is small, so that the
differences in observed mass function ($f$) are mainly due to
differences in inclination. For those systems with inclinations viewed
close to edge-on ($i \sim 90^\circ$), we detect eclipses;
those with smaller inclinations have mass functions that are lower
by a factor of $\sin^3 i$ and no eclipses are detected.
To test this hypothesis, it is important to check that the
``non-eclipsing'' VLMBPs like PSR~A do indeed lack
eclipses. In \S 2 we describe multi-frequency observations
with the Green Bank and Arecibo radio telescopes. In \S 3, we describe
the search for new pulsars in the Arecibo data, and present the new
pulsar discovery, PSR~J1911+0101B (henceforth PSR~B). We also present
the timing
solutions of both pulsars, and describe how they were obtained. The
properties of these timing solutions are discussed in \S 4; these can
be used to search for eclipses of PSR~A near superior conjunction, to
check the validity of the mass model of NGC~6760, and to
probe the interstellar medium along the line of sight. Finally, in \S
5, we conclude with a summary of the main results and prospects for
future observations.

\section{Observations and their motivation}

Observations were made with both the 100-m Robert C. Byrd Green Bank
Telescope (GBT) and the 305-m Arecibo radio telescope.  The GBT can
observe in the important 500 and 800\,MHz bands and can track
PSR~A over a full orbit. Arecibo provides superior gain for
observations near 1400\,MHz.

\subsection{GBT Observations}

The properties of eclipses in VLMBPs are highly frequency dependent.
As a rule, eclipses become more dramatic as the frequency decreases.
For example, the eclipse lengths, $\bigtriangleup T$, of PSRs
B1744$-$24A, B1957+20 and J2140$-$2310A
scale inversely with frequency, $\nu$, as $\bigtriangleup
T\,\propto\,\nu^{\beta}$, where $\beta\,=\,-0.63 \pm 0.18$,
$-0.41 \pm 0.09$ and $-0.45 \pm 0.05$
respectively \cite{nttf90,fbb+90,rsb+04}. Since eclipses
observed at low frequencies are extended along the orbital plane, it
is reasonable to suppose that they are extended off the orbital plane as
well, so that, for lines-of-sight far from the plane (i.e., at low
inclinations), eclipses might be detectable only at low frequencies.
Further evidence that eclipses are more detectable at lower frequencies
comes from PSRs J2051$-$0827 and J0023$-$7203J.  The former shows no change
in flux density and only small dispersive delays at 1400 MHz, while
it is partially or completely eclipsed near 600 MHz (Stappers et
al. 1996, Nice, Stairs and Arzoumanian 2004)\nocite{sbl+96,nsa04}.
The latter shows only small dispersive delays at
1400 MHz and occasional eclipses at 660 MHz, while eclipses occur
regularly at 430 MHz \cite{fck+03}. Given these properties, we made
low-frequency observations of PSR~A with the GBT at several
frequencies between 575 and 1660\,MHz (see also Nice, Stairs and
Arzoumanian, 2004\nocite{nsa04}, which also includes results for other VLMBPs).
Data were collected by the Spectral Processor, a Fourier transform
spectrometer that allows a total bandwidth of 40~MHz. Spectra were
folded on-line modulo the pulse period over intervals of 5
minutes. For these observations, we used two different modes: 512
channels across 40 MHz, folded into 94-bin profiles (July 2002, and in
some scans taken in October 2002) and 1024 channels across 40 MHz,
folded into 70-bin profiles for the remaining data (taken in October
2002). Off-line, the folded spectra were de-dispersed, and pulse times
of arrival (TOAs) were extracted using conventional techniques (see,
for example, Taylor 1992).\nocite{tay92}

\subsection{Arecibo Observations}

Dispersion measure variations can be very effectively probed at higher
frequencies using a telescope with a large gain and observing
bandwidth. For this reason, and also to search for new pulsars and
determine phase-connected timing solutions, we have been observing
NGC~6760 since March 2003 with the Arecibo telescope. A total of
eighteen observations were made, lasting as long as 4500\,s, using the
new L-band wide receiver. This system 
provides unparalleled sensitivity from 1100 to
1730\,MHz where its equivalent flux density is only 2.5\,Jy. For data
acquisition, we used three of the four\footnote{The fourth WAPP was not
used for these experiments due to the high levels of radio frequency
interference in the remaining part of the band.} Wideband Arecibo
Pulsar Processors (or WAPPs, Dowd, Sisk \& Hagen 2000)\nocite{dsh00},
with each WAPP providing 512 channels covering 100-MHz spectral
windows centered at 1170, 1410 and 1510~MHz respectively. Some of the
early observations used one or two WAPPs only. At the DM of
PSR~A, the dispersion smearing in each frequency channel
ranges between 94$\,\mu$s at 1510~MHz and 202$\,\mu$s at 1170~MHz. The
sampling time was 128$\,\mu$s, so that the corresponding time
resolution ranged between 160 and 240$\,\mu$s.
To extend the timing baselines, we also used data from a previous
Arecibo survey, taken June 23 to 30, 2001, which targeted NGC~6760 using a
different receiver and a single WAPP centered at 1525~MHz
\cite{rhs+04,hrs+03}.

\section{Results}

\subsection{Pulsar search}

Assuming that the root mean square (rms) the off-pulse samples
is the noise level predicted by the radiometer equation for the 
the sky temperature for the location of NGC~6760 at 1400 MHz
($\sim$4.6~K),
we estimate that the flux density of PSR~A at 1410 MHz is about
0.2 mJy. This value is substantially different from the
estimate of Deich~et~al.~(1993) of 1.5~mJy, but we believe it to be
essentially correct. We estimate that the associated
uncertainty on this value is about 30\%. This pulsar is detected with
an average signal-to-noise ratio (S/N) of about 200 in our 4500-s
Arecibo observations. Motivated by the high sensitivity of the
Arecibo observing system, we carried out a search for additional 
pulsars in NGC~6760 using the Borg, a 104-processor Beowulf
cluster at McGill University, running the {\tt PRESTO} software
package \cite{ran01} in a similar fashion to that described by
Ransom et al.~(2004b).\nocite{rsb+04} In brief, the accumulated
correlation functions recorded by each WAPP were transformed into the
equivalent spectral frequency channels and then dedispersed. We
searched 41 trial DM values in the range 190-210 cm$^{-3}$~pc, in
steps of 0.5~cm$^{-3}$~pc. For each trial DM value, a full
Fourier-domain matched-filter acceleration search was carried out
(Ransom, Eikenberry and Middleditch 2002)\nocite{rem02}, enabling the
detection of binary systems with relatively short orbital periods.
The maximum detectable orbital acceleration as a function of spin
period is:
\begin{equation}
a_{\rm max} = \frac{Z_{\rm max} c P}{T^2_{\rm obs}},
\end{equation}
where $Z_{\rm max}$ is the maximum number of spectral bins that the
fundamental harmonic can drift during the whole observation time
$T_{\rm obs}$. In our searches, $Z_{\rm max} = 500$ and $T_{\rm obs} \sim
4500$s. For a 1-ms pulsar, this represents an acceleration limit of
7.4 m\, s$^{-2}$. In order to detect highly accelerated binary pulsars,
we have also searched short, overlapping sections of the observations.
This search procedure was applied to a total of ten observations.

\begin{figure}
\setlength{\unitlength}{1in}

\begin{picture}(0,3.0)
\put(0.0,4.8){\includegraphics{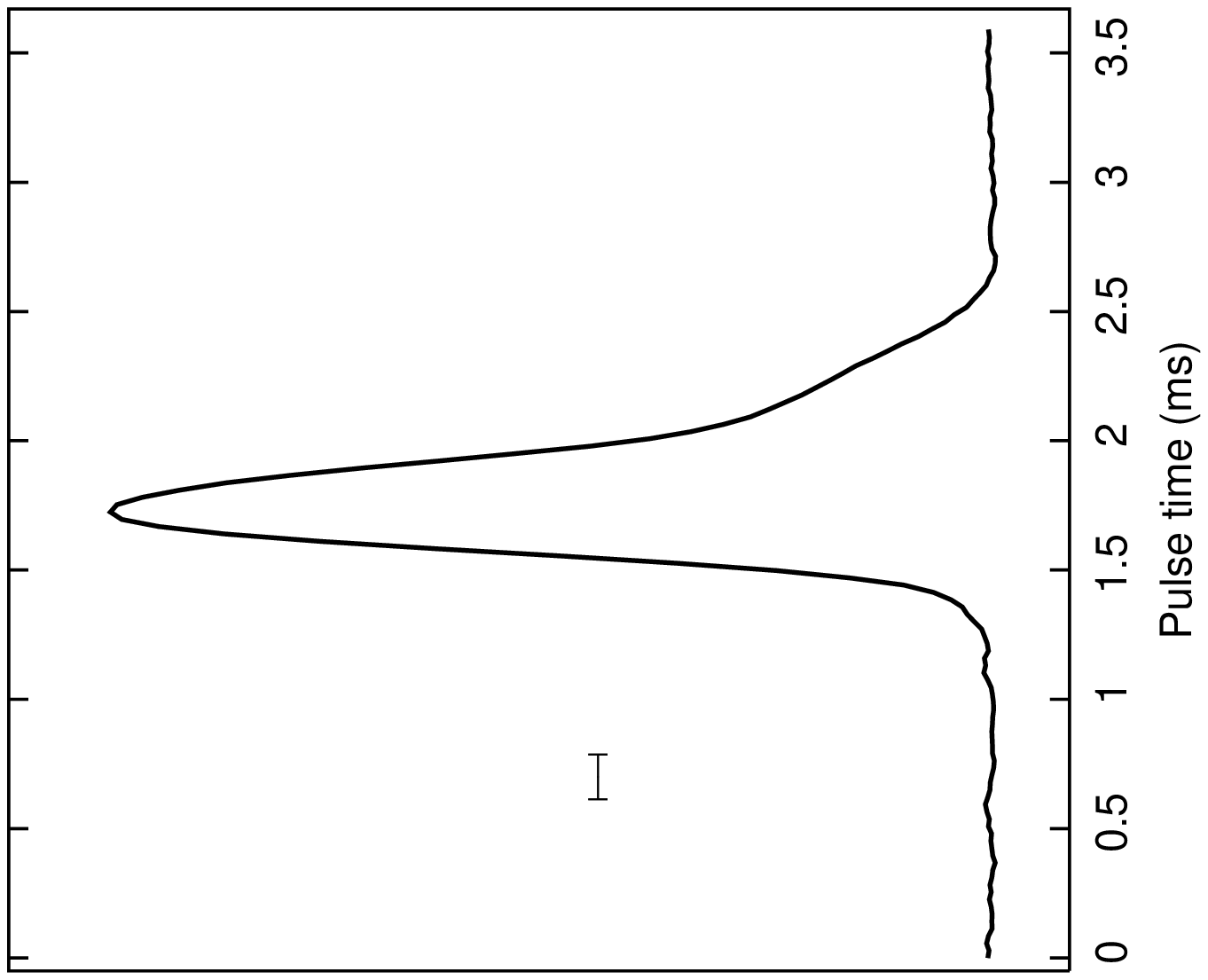}}
\put(3.0,4.8){\includegraphics{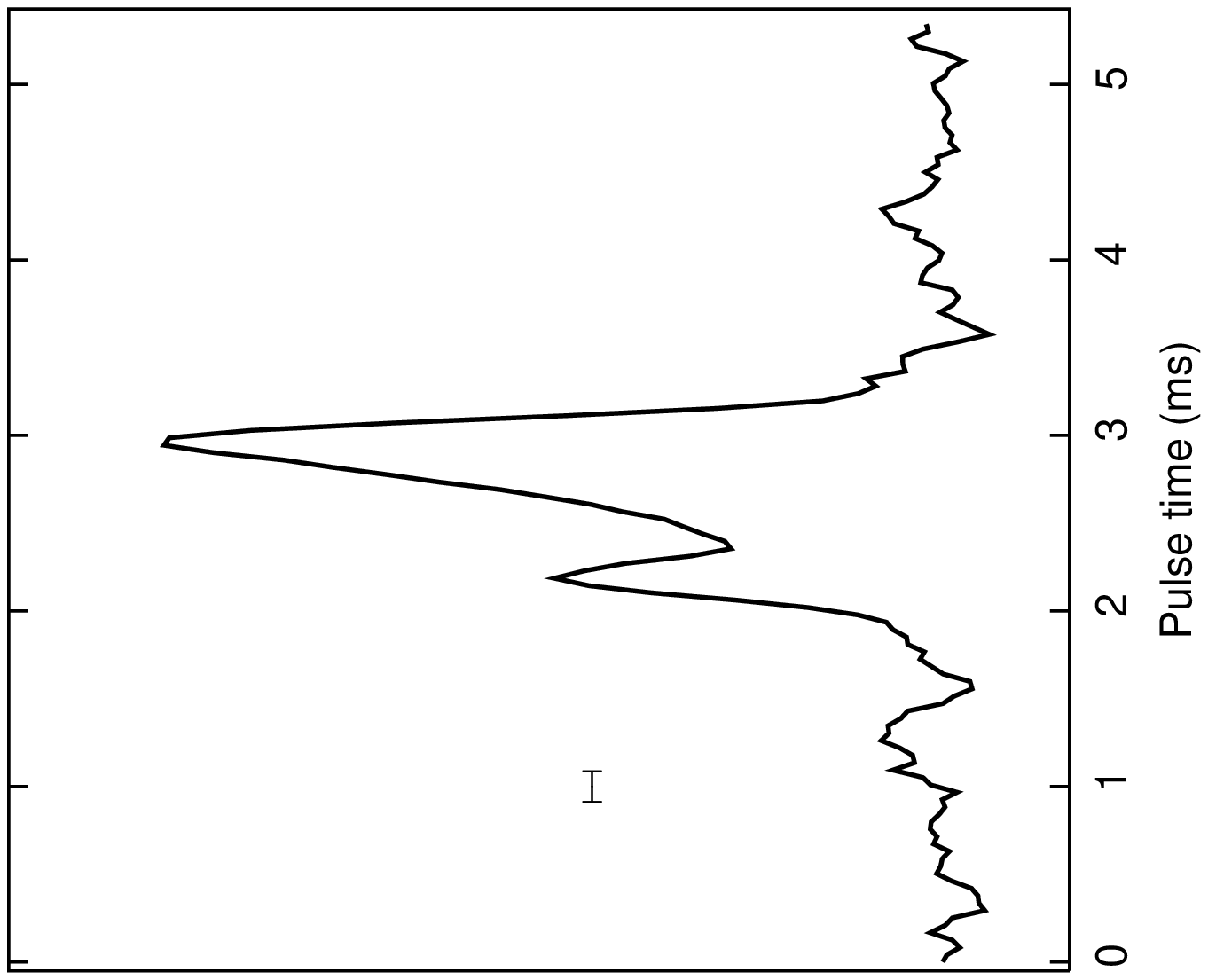}}
\end{picture}
\caption [] {\label{fig:profiles}
Full-cycle pulse profiles for the two known pulsars in NGC~6760. These were obtained 
by averaging all the Arecibo WAPP data from the bands centered at 1410 and 1510 MHz. The
horizontal error bars indicate the instrumental time resolution at 1410 MHz.
Left: PSR~A, right: PSR~B.}
\end{figure}

No new accelerated pulsars were found.  However, one new isolated
5.38-ms pulsar (PSR~B) was found at a DM of 196.7~cm$^{-3}$~pc as a
result of this processing. It is detectable with a S/N of about 10 in
an hour. This pulsar is about 8 times fainter than
PSR~A, i.e., its flux density  at 1410 MHz is $\sim\,0.026\,$mJy, its
luminosity is about 2 mJy kpc$^2$. The pulse profiles for PSR~A and
PSR~B can be seen in Fig.~\ref{fig:profiles}. Interstellar scattering
is not detectable, as expected: the Cordes and Lazio NE2001 model
\cite{cl02} predicts a scattering timescale of only 0.04 ms at 1400~MHz.

\subsection{Pulsar Timing}

\begin{figure}
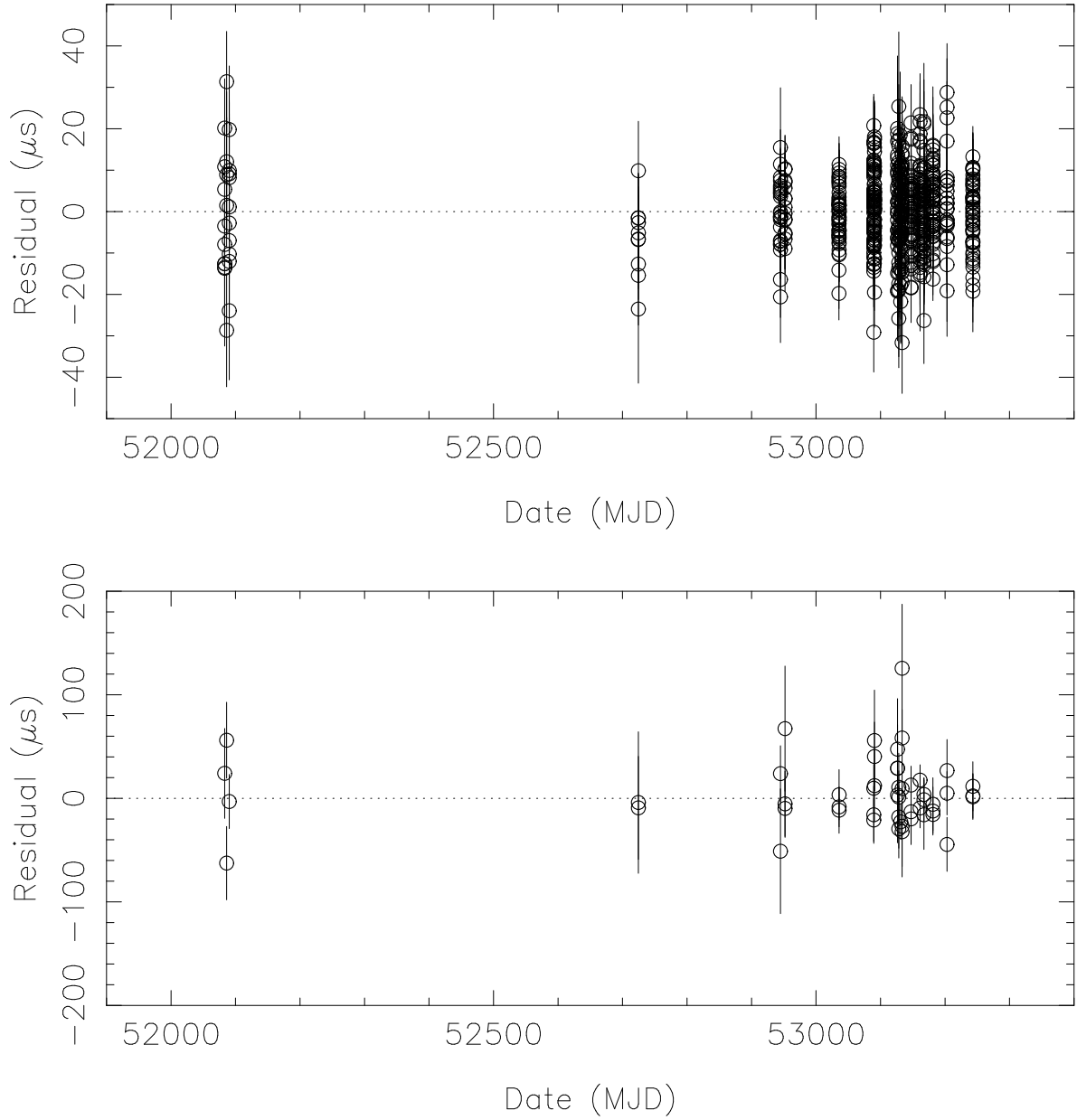

\setlength{\unitlength}{1in}
 
\begin{picture}(0,6.0)
\put(-0.7,9.6){\includegraphics{./fig2a.ps}}
\put(-0.7,6.2){\includegraphics{./fig2b.ps}}
\end{picture}
\caption [] {\label{fig:residuals}
Post-fit Arecibo residuals for PSR~A (top) and PSR~B
(bottom). Notice the difference in vertical scale. The early Arecibo
search data is at the left, near MJD\,=\,52000.
}
\end{figure}

We have used our set of observations to obtain phase-coherent
timing solutions for both of the pulsars now known in NGC 6760.
After the pulse periods and DMs were established by the search
analysis, we used the {\tt SIGPROC} package \footnote{see
http://www.jb.man.ac.uk/$\sim$drl/sigproc/} \cite{lor01b} to
dedisperse the data and the {\tt PRESTO} software package to fold
the resulting time series and to extract pulse times of arrival
(TOAs). These TOAs were then analyzed with
{\tt TEMPO}\footnote{http://pulsar.princeton.edu/tempo/}. 
We used the JPL's DE/LE 200 solar system ephemeris to subtract the
effect of the motion of the telescope relative to the barycenter of
the solar system. As is now standard practice for low-eccentricity
binary systems like PSR~A, the ELL1 binary model \cite{lcw+01} was used in
{\tt TEMPO} in the fitting process. The resulting parameters are presented
in Table \ref{tab:solutions}; the TOA residuals are displayed in
Fig.~\ref{fig:residuals}, they display no obvious trends. This effort
was greatly aided by the
conspicuous lack of scintillation effects, as expected for pulsars
with large DMs. Phase connection between 2001 and 2003 was confirmed
using the GBT data collected in 2002. In particular, a well-known 1\,s
instrumental offset in the 2001 data, detected by other WAPP timing
projects \cite{rhs+04} was confirmed and removed. Once phase
connection was achieved, the GBT data were removed from subsequent
analysis; due to their relatively low signal-to-noise ratios and high
measurement uncertainties, they added little to the timing
solution. The number of rotations of PSR~A and PSR~B between the 2001 and 2003
observations is unambiguous.

\section{Discussion}

\subsection{Search for Eclipses}

\begin{figure}
\setlength{\unitlength}{1in}

\begin{picture}(0,6.5)
\put(0,-0.2){\includegraphics{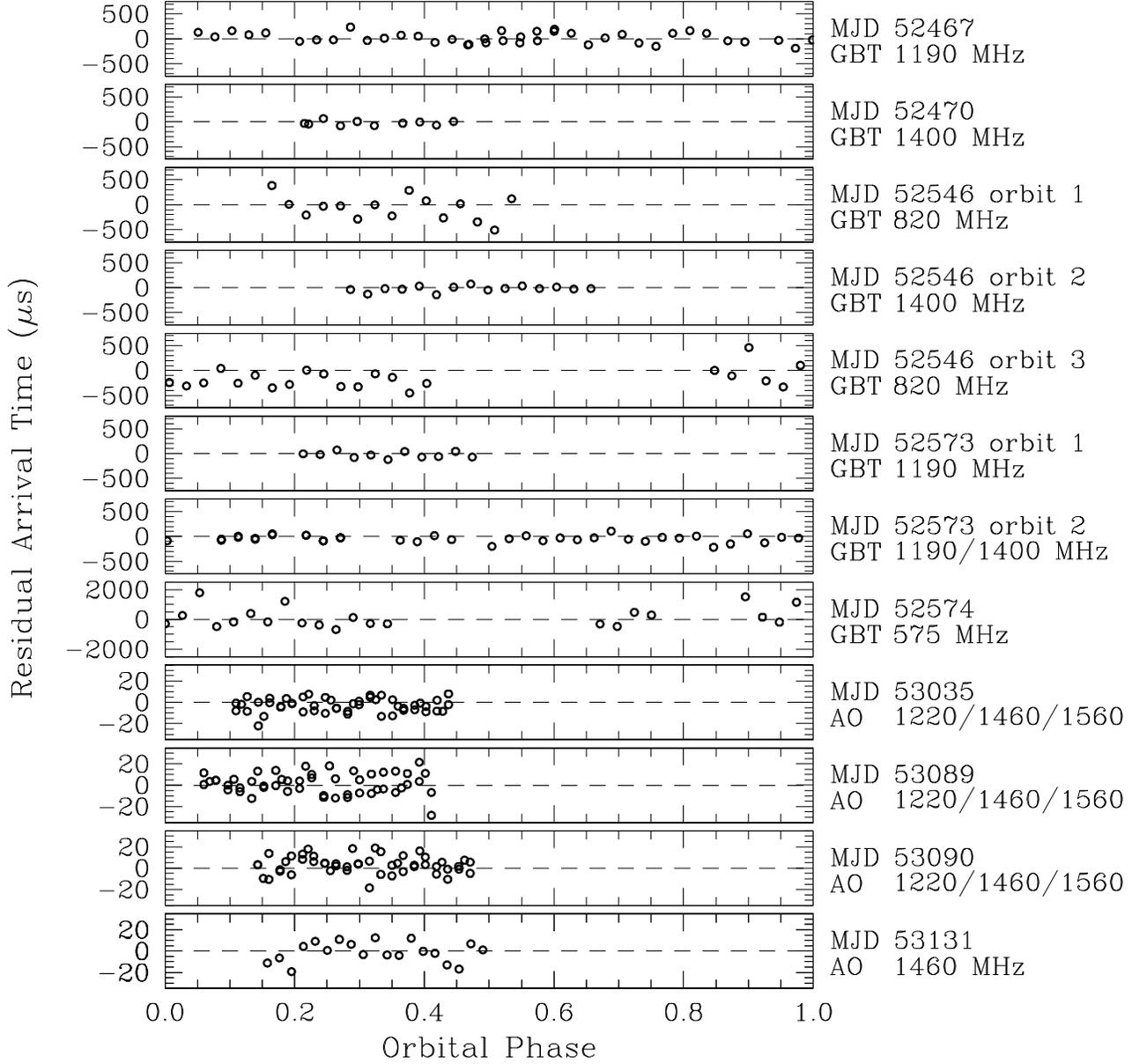}}
\end{picture}
\caption [] {\label{fig:orbital}
Residual pulse arrival times of PSR~A, measured at the GBT
and Arecibo over 12 orbits on nine separate days. Observing
frequencies and telescope used are indicated in the
figure. A partial eclipse would be indicated by arrival time delays ,
or complete lack of signal, at orbital phase 0.25. {\it All}
observations on the indicated days are shown in the figure, there are
{\it no} instances of missing signals. Gaps in some orbits indicate
times at which data were not collected, not an intrinsic fading of the
pulsed signal. At Arecibo the partial coverage is caused by the
limited field of view of the telescope. There are other observations
made at Arecibo, which are not displayed because they do not cover the
relevant orbital phase.
}
\end{figure}

We found no evidence for eclipses of PSR~A, as those would
have been detected by either the absence or systematic reduction in
flux density around superior conjunction or, in the case of partial
eclipses, systematic delays of the pulsed signal as it passed
through the ionized eclipsing medium. Neither of these observational
signatures were detected. The data are shown in Fig.~\ref{fig:orbital}. 
Residual pulse arrival times are shown after
subtracting the best available orbital model, presented in Table
\ref{tab:solutions}. The ascending node is at orbital phase $\phi = 0$, 
and superior conjunction and any eclipses should occur at $\phi =
0.25$.

Limits on the electron column density of ionized eclipsing
material can be derived from limits on the TOA delays, $\bigtriangleup
t$ at $\phi = 0.25$.
We find $\bigtriangleup t \leq 200 \mu$s at 820~MHz (GBT) and
$\bigtriangleup t \leq 5 \mu$s at 1120-1560~MHz (Arecibo), yielding
upper limits of $\bigtriangleup \rm DM\,<\,0.03\,pc\,cm^{-3}\,=\,
10^{17}\,cm^{-2}$ and $\bigtriangleup\rm\,DM\,<\,0.0017\,pc\,cm^{-3}\,\sim
5\,\times\,10^{15}~cm^{-2}$ respectively.  The latter limit is well under
the observed dispersion delays in PSR J2051$-$0827 \cite{sbl+96,sbl+01}
and PSR J0023-7203J \cite{fck+03}, for which $\bigtriangleup \rm
DM\,\sim\,3\,\times\,10^{17}$ and
$\sim\,2\,\times\,10^{16}\,\rm cm^{-2}$ respectively.

Observations below 800 MHz could put more stringent limits
on dispersive delays. However, we found the pulsar very difficult to
detect at lower frequencies. It was only marginally detectable in
5-minute GBT observations at 575 MHz, as can be inferred by the large
scatter in the timing residuals on MJD 52574 in Figure
\ref{fig:orbital} (Because those residuals tend to cluster around
zero, we are confident that the pulsar was, in fact, being detected.)

The lack of eclipses provides for circumstantial evidence that the
system's small mass function (lower than all of the eclipsing
VLMBPs) is indeed due to a low inclination angle.

\subsection{Positions and Accelerations}

The best-fit Right Ascension ($\alpha$) and Declination ($\delta$) of
PSR~A and PSR~B indicate that these pulsars are very close to the
center of the cluster. The projected angular distances from the center
are indicated in Table \ref{tab:solutions} as $\theta_{\perp}$, these
are equivalent to 1.25 and 0.35 core radii. In this respect, these
pulsars are similar to the majority of the pulsars known in GCs.

The observed period derivatives ($\dot{P}_{\rm obs}$) of both pulsars
are negative. As discussed in detail by Damour \& Taylor (1991)\nocite{dt91},
the measured period derivative for a pulsar is a sum of several terms:
\begin{equation}
\label{eq:pdot}
\left( \frac{\dot{P}}{P} \right)_{\rm obs} = \left( \frac{\dot{P}}{P}
\right)_{\rm int} + \frac{\mu^2 D}{c} + \frac{a_z}{c}.
\end{equation}
The first term on the right is the pulsar's intrinsic period
derivative ($\dot{P}_{\rm int}$) divided by its rotational period
($P$), the second term is a fictitious centrifugal acceleration (known as the
Shklovskii effect), where $\mu$ is the proper motion of the pulsar
(expected to be close to the mean proper motion of the cluster),
and the third term is the pulsar's line-of-sight acceleration. The first
two terms on the right are essentially unknown (no proper motion has
yet been measured for NGC~6760), but are always positive. We
therefore know that the pulsar's line-of-sight acceleration relative
to the solar system barycenter, $a_z/c$ must be smaller than
$(\dot{P}/P)_{\rm obs}$.

The line-of-sight acceleration term has a contribution from the
acceleration of NGC~6760 itself in the gravitational field of the
Galaxy relative to the solar system barycenter. This can be estimated
using a mass model for the Galaxy: $a_{z\rm
  G}\,=\,-2.78\,\times\,10^{-10}\rm m\, s^{-2}$ for $D\,=\,9.5\,\rm
kpc$ \cite{pac90}. Subtracting this term from $(\dot{P}/P)_{\rm obs}$,
we obtain upper limits for the line-of-sight component of the
gravitational field of NGC~6760 at the locations of the pulsars; this
is indicated in Table \ref{tab:solutions} as $a_{z \rm C}$. The fact
that this quantity is negative for PSR~A implies that it is on the far
side of NGC~6760.

To check whether these values are reasonable, we implemented a simple
cluster mass model (see Appendix 1 for details) to calculate the
acceleration along the line of sight as a function of the pulsar's
projected distance from the center of the cluster ($r_{\perp}\,\equiv\,
D \theta_{\perp}$), the cluster's core radius ($r_c\,=\,\theta_c D$)
and rms of spectroscopic stellar velocities at $\theta_{\perp}=0$,
$\sigma_z(0)$. For
NGC~6760, $\sigma_z(0)=5.77$~km~s$^{-1}$ \cite{web85}. The minimum
acceleration of PSR~A can be comfortably accounted for by this model,
which predicts maximum line-of-sight accelerations of 1.14 and
1.75$\,\times\,10^{-9}\,\rm m\,s^{-2}$ at the locations of PSR~A and
PSR~B.
If we assume that the pulsars cannot have an acceleration larger than
this value, then from equation \ref{eq:pdot} we can derive upper
limits for the intrinsic period derivatives. From these, we can derive
lower limits for the characteristic ages ($\tau_c$) and upper limits
for the magnetic flux densities at the magnetic poles ($B_0$); all
these limits are presented in Table \ref{tab:solutions}. Both pulsars
are quite old, and the magnetic flux densities at their poles must be
relatively small. Such strong constraints can only be placed because
of the small accelerations predicted by the NGC~6760 mass model.
For most of the allowed range of $\dot{P}_{\rm int}$ for PSR~B, the
pulsar has a negative acceleration, indicating that it is more likely
on the far side of the cluster.

\subsection{DM Spread}

\begin{figure}
\setlength{\unitlength}{1in}
 
\begin{picture}(0,6.0)
\put(0.4,0){\includegraphics{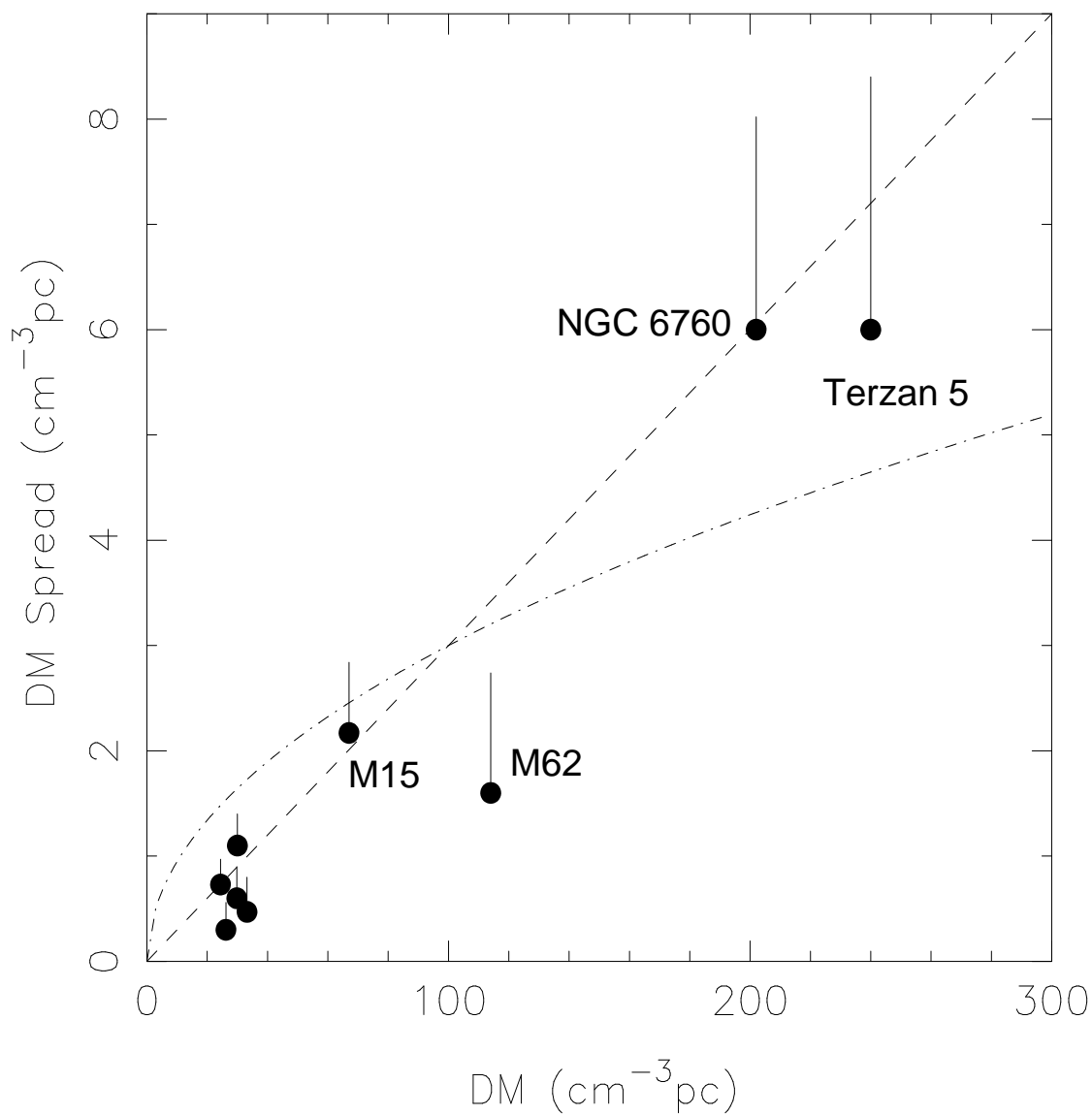}}
\end{picture}
\caption [] {\label{fig:DM_scatter}
Observed DM spread for the pulsars in GCs as a function of
DM. These observed values are always lower limits;
new pulsar discoveries in each GC can only increase spread in the
future, this is symbolically indicated by the vertical line segments
(their length is arbitrary, but proportional to the DM of the cluster).
For meaningful DM spread estimates, only GCs with three or more
known pulsars were included, with the exception of NGC~6760: for this
object, the measured DM spread is already quite significant, despite
the fact that only two pulsars are known. The dashed line indicates
than no known cluster has shown a DM spread significantly 
larger than 3\% of its DM. The curved dot-dashed line indicates the best
fit for a DM spread proportional to DM$^{0.5}$; this is clearly a much
poorer fit.
}
\end{figure}

The spread of DMs in NGC~6760 (196.68 to 202.68~cm$^{-3}$pc) is
remarkably large, about the same as
Terzan~5 \cite{lmbm00,ran01}. The origin of the DM spreads is
an interesting question. For the pulsars in
47~Tucanae, Freire~et~al.~(2001)\nocite{fkl+01} showed that the DM
variation in that GC is mostly due to the cluster gas itself. For M15,
with a higher central DM; the DM variation between pulsars has been
attributed to sub-arcminute scale irregularities in the Galactic
electron column density as a function of sky position
\cite{and92}. For both Terzan~5 and NGC~6760, the spreads of pulsar DMs
are much larger still, despite the large differences in the nature of
the clusters. This indicates that the spread of DMs in Terzan~5 is
probably not due to the presence of gas in the cluster, but instead to
a cause common to both clusters. For high electron column densities, one
would generally expect that variations of the DM with sky position at
these sub-arcminute scales will play a larger role. The observed
correlation between DM and $\Delta$DM is approximately linear (see
Fig. \ref{fig:DM_scatter}), indicating that it will be very
difficult to detect the plasma content of GCs at high DMs. 
This result
is also important for GC pulsar searches since it provides a guideline
for the range of trial DMs required to be sensitive to pulsars in a 
given cluster. 

If the plasma irregularities in space were completely random we would
expect a random-walk-like dependence of $\Delta$DM versus DM, i.e.,
the spread in DM would be proportional to the square root of the
DM. This is, however, not consistent with the observations, which
indicate instead a linear relation of DM variance and DM. According to
Backer~et~al.~(1993)\nocite{bhh93} and references therein, this
implies that the ISM is turbulent in nature at the spatial scales
being probed by pulsar separations in globular clusters, about 1 pc.

\section{Conclusions and prospects}

In a comprehensive study of PSR~J1911+0102A, we have failed to detect
any eclipses, but we have determined this pulsar's timing solution. This,
together with the low mass function of the system, adds weight to the
hypothesis that the observational properties of VLMBPs are mainly
determined by the inclination of the system. Timing studies at Arecibo
show the pulsar to be located near the center but on the far side of
NGC~6760, and to have a large characteristic age and small
magnetic field. These data further constrain the occurrence of partial
eclipses. A second isolated 5.38-ms pulsar PSR~J1911+0101B was
discovered in these data. It too is located near the center of the
cluster and probably on its far side. The large difference in DM
between the pulsars suggests that the ISM is turbulent on spatial
scales similar to their separation, about 1 pc.

By providing precise positions for the pulsars, the timing solutions
make it possible to identify these pulsars at optical/near-infrared and
X-ray wavelengths. Continued timing
will make it possible to measure the proper motion of
NGC~6760, to search for expected orbital variability of PSR~A (as observed for
other VLMBPs, see references in Freire, 2004)\nocite{fre04}, to
monitor possible DM variability and to check for the presence of
distant stellar or planetary companions to either PSR~A or PSR~B.

\acknowledgements

The Arecibo Observatory, a facility of the National Astronomy and
Ionosphere Center, is operated by Cornell University under a
cooperative agreement with the National Science Foundation.
The National Radio Astronomy Observatory is a facility of the National
Science Foundation operated under cooperative agreement by Associated
Universities, Inc. JWTH is an NSERC PGS-D fellow. DN acknowledges
support from NSF grant 0206205. DRL is
a University Research Fellow funded by the Royal Society. IHS
holds an NSERC University Faculty Award and is supported by a
Discovery Grant.  The computing facility used for this research was
funded via a New Opportunities Research Grant from the Canada
Foundation for Innovation. We also thank Prof. I. King for his
suggestions, which led to the implementation of the globular cluster
model described in the Appendix, and C. Salter for a first skeptical
review. We are greatly indebted to Jeffrey R. Hagen and William Sisk
of NAIC for their excellent work developing the WAPP, a world-class
research instrument.

\section*{Appendix: Cluster Model}

We implemented a very simple analytical model of NGC~6760
based on the empirical density law for dense star clusters derived by
King~(1962)\nocite{king62}. For a star cluster with an angular tidal
radius that is significantly larger than the core radius, the surface
mass density near the center is given by:
\begin{equation}
\label{eq:surface_density}
f(r_{\perp}) = \frac{f_0}{1 + (r_{\perp}/r_c)^2},
\end{equation}
where $f_0$ is the central surface mass density.
The corresponding spatial density is:
\begin{equation}
\label{eq:space_density}
\rho(r) = \frac{\rho_0}{[1 + (r/r_c)^2]^{3/2}},
\end{equation}
where $r$ is the linear distance to the center of the cluster,
defined as $r^2 = z^2 + r_{\perp}^2$, where $z$ is the distance
of any point to the plane of the sky passing through the center of the
cluster, and $\rho_0 = f_0 / 2 r_c$ is the central density.
 Equations \ref{eq:surface_density} and
\ref{eq:space_density} correspond to the limits of eqs. 14 and 27 of
King~(1962) for infinite tidal radii, and do not apply to core-collapsed
clusters.
We can integrate eq. \ref{eq:space_density} radially to obtain the
mass inside a given radius $r$:
\begin{eqnarray}
\label{eq:mass}
M(r) & = & \int_{0}^{r} \rho (r') 4 \pi r'^2 dr'\\
     & = & 4 \pi \rho_0 r_c^3
\left[\sinh^{-1}\left(\frac{r}{r_c}\right) - \frac{r}{r_c \sqrt{1 + (r/r_c)^2}}\right] \nonumber,
\end{eqnarray}
where $\sinh^{-1}x$ can also be expressed as $ \ln \left( x + \sqrt{1 +
  x^2} \right)$. Therefore, eq. \ref{eq:mass} diverges logarithmically
for infinite radii, but it is a good approximation near the center of the
cluster.

The acceleration at $r$, $a(r)$, can be obtained by multiplying
$M(r)$ by $-G/r^2$. From Spitzer (1997)\nocite{spi87}, we find that $\rho_0 = 9
\sigma^2_z(0)/(4 \pi G r^2_c)$. This allows us to present $a(r)$
as a function of well-known observables:
\begin{equation}
\label{eq:accel2}
a(r) = \frac{9 \sigma^2_z(0)}{D \theta_c} \frac{1}{x^2}
\left( \frac{x}{\sqrt{1+x^2}}  - \sinh^{-1}x \right),
\end{equation}
where $x \equiv r / r_c$. The line-of-sight component $a_z(r)$ can
be obtained by multiplying $a(r)$ by $z/r$. For each of the pulsar
line-of-sight distances to the center of the cluster, $r_{\perp}$, we
calculate $a_z(r)$ for a variety of values of $z$, recording the
maximum values found.

Integrating eq. \ref{eq:accel2} from the center of the cluster until a
given location $r$, one obtains the gravitational potential relative to the
center of the cluster:
\begin{eqnarray}
V - V_0 & = & - \int_{0}^{r} a(r') dr' \\ \nonumber
& = & - 9 \sigma^2_z(0) \int_{0}^{x}\frac{1}{x'^2}
\left[ \frac{x'}{\sqrt{1+x'^2}}  - \ln \left( x' + \sqrt{1 +  x'^2} \right) \right]
dx'\\ \nonumber
& = & 9 \sigma^2_z(0) \left[ 1 - \frac{\ln \left( x + \sqrt{1 +
      x^2} \right) }{x} \right],
\end{eqnarray}
where $x' \equiv r'/r_c$.

\begin{deluxetable}{lcc}
\tabletypesize{\footnotesize}
\tablecolumns{3}
\tablewidth{0pc}
\tablecaption{PARAMETERS FOR THE NGC 6760 PULSARS}
\tablehead{ \colhead{Pulsar} & \colhead{PSR J1911$-$0102A} & \colhead{PSR
  J1911$-$0101B}}
\startdata
\multicolumn{3}{c}{Observation and data reduction parameters}\\
\hline
Period Epoch (MJD)    & 53000 & 53000 \\
Start time (MJD)     & 52083 & 52083 \\
End time (MJD)       & 53243 & 53243 \\
\# of TOAs  & 624 & 48 \\
TOA rms ($\mu$s) & 8.4 & 22 \\
\hline
\multicolumn{3}{c}{Timing parameters}\\
\hline
$\alpha$ \tablenotemark{a}& $19^{\rm h}11^{\rm m}11\fs 08957(4)$ & $19^{\rm h}11^{\rm
  m}12\fs 5725(4)$ \\
$\delta$ & $+01^\circ 02\arcmin 09\farcs 741(2)$ & $+01^\circ 01\arcmin
50\farcs 44(2)$ \\
$P$ (ms)        & 3.618524251059(2) & 5.384325706188(9) \\
$\dot{P}_{\rm obs}$ (10$^{-21}$) & $-$6.58(2) & $-$2.0(3)\\
DM ($\rm cm^{-3}\,pc$) & 202.678(3) &  196.69(2)\\
$P_b$\tablenotemark{b}(days)    & 0.1409967943(6) & - \\
$T_{\rm asc}$ (MJD) & 53000.0517850(8) & - \\
$x$ (s)         & 0.037658(2) & - \\
$e$             & $<$0.00013 & - \\
\hline
\multicolumn{3}{c}{Derived parameters}\\
\hline
$\theta_{\perp}$ (\arcmin) & 0.42 & 0.12 \\
$a_{z \rm C}$ (m\,s$^{-2}$) & $<\,-2.6\,\times\,10^{-10}$ & $<+\,1.6\,\times\,10^{-10}$ \\
$\dot{P}_{\rm int}$ & $<\,1.06\,\times\,10^{-20}$ &
$<\,3.35\,\times\,10^{-20}$ \\
$\tau_c$ (Gyr) & $>\,5.4$ & $>\,2.5$\\
$B_0$ (gauss) & $<\,2.0\,\times\,10^8$ & $<\,4.3\,\times\,10^8$ \\
$f\,(\rm M_{\odot})$ & 0.0000028842(4) & - \\
$m_c$\tablenotemark{c}$(\rm M_{\odot})$ & $\sim$0.02 & - \\
\enddata
\tablenotetext{a}{The uncertainties indicated for all parameters
are 1-$\sigma$, and are twice the formal values given by {\tt TEMPO}.}
\tablenotetext{b}{The orbital parameters are: orbital
period ($P_b$), time of passage through the ascending node ($T_{\rm
  asc}$), semi-major axis of the orbit of the pulsar, projected along
the line-of-sight, divided by the speed of light ($x$) and orbital
eccentricity ($e$). Since the latter quantity is too small to be
measured, we can not estimate the longitude of the periastron relative
to ascending node ($\omega$). All other parameters are as described in
the text.}
\tablenotetext{c}{To calculate the companion mass $m_c$, we assumed an
inclination angle $i=60^\circ$ and a canonical pulsar mass of
1.4~M$_{\odot}$.}
\label{tab:solutions}
\end{deluxetable}

\end{document}